\documentstyle[12pt]{article}
\begin{document}
\begin{titlepage}
\title{ISOCURVATURE PERTURBATIONS IN MULTIPLE INFLATIONARY MODELS}
\author{David POLARSKI$^{1,2}$ and A.~A.~STAROBINSKY$^{3,4}$ \null \\
$^1$Lab. de Mod\`eles de Physique Math\'ematique, EP93 CNRS\\
Univ. de Tours, Parc de Grandmont, F-37200 Tours (France)\\
$^2$D\'epartement d'Astrophysique Relativiste et de Cosmologie,\\
Obs. de Paris-Meudon, UPR 176 CNRS, 92195 Meudon cedex (France)\\
$^3$Yukawa Inst. for Theor. Physics, Kyoto Univ., Uji 611 (Japan)\\
$^4$Landau Institute for Theoretical Physics,\\
Kosygina St. 2, Moscow 117334 (Russia)\\}
\date{\today}
\maketitle
\begin{abstract}

Dynamics of long-wave isocurvature perturbations during an inflationary
stage in multiple (multi-component) inflationary models is calculated
analytically for the case where
scalar fields producing this stage interact between
themselves through gravity only. This enables to determine
correct amplitudes of such perturbations
produced by vacuum quantum fluctuations of the scalar fields during
the multiple inflationary stage. Exact matching
to a post-inflationary evolution that gives the amplitude of
isocurvature perturbations in the cold dark matter model with
radiation is performed in the case where a massive inflaton field remains
uncoupled from usual matter up to the present time. For this model,
isocurvature perturbations are smaller than adiabatic ones in the region
of the break in the perturbation spectrum which arises due to a
transition between the two phases of inflation, but they may be much
bigger and have a maximum at much shorter scales. The case of an inflaton
with a quartic coupling which remains uncoupled after inflation is
considered, too.

\end{abstract}
\end{titlepage}

\section{Introduction}

Inflationary cosmological models in which a de Sitter (inflationary)
stage is produced by a number of effective scalar fields (inflatons)
are called multiple (or multi-component)~\cite{star85}
(see~\cite{linde} for a general review). A double inflationary
model with two scalar fields~\cite{linde85}-\cite{david} is a specific
case of them. Note
that extended inflationary~\cite{la89} models or inflationary models
in the Brans-Dicke theory of gravity may also be considered as
belonging to this class of models after
transformation to the Einstein frame. Double inflationary models
producing a step-like spectrum of initial adiabatic perturbations
give a possibility to reconcile the CDM model with observations
without introducing neutrinos ~\cite{muc,ppst}.
If $N$ is the number of light scalar fields at the inflationary stage
in such a theory ($|m_i^2|\ll H^2,~ H\equiv \dot a/a$,
where $a(t)$ is the scale factor of the FRW isotropic cosmologiocal
model), then $N$ independent branches of non-decaying quantum
fluctuations of the
scalar fields are generated during the inflationary stage similar, and in
addition, to quantum fluctuations of gravitons (the resulting energy
spectrum of the latter was first correctly calculated in ~\cite{star79}).
However, only one linear combination of these fluctuations produces
the growing scalar (adiabatic) mode which is usually assumed to be
responsible for the formation of galaxies, stars (and other compact
objects) and the large-scale structure of the Universe.
The other $N-1$ modes are isocurvature fluctuations during the
inflationary stage (they were first considered in ~\cite{lin84}).

Isocurvature fluctuations are less universal than adiabatic ones.
First, they only appear in multiple, not single inflationary models.
Second,
they might not survive up to the present time (and typically do not).
Really, they can exist now only if at least one of the inflaton scalar
fields remains non-thermalized and uncoupled from the usual matter
(radiation, baryons and leptons) during the whole evolution of the
Universe from the inflationary era until the present period (so that the
corresponding particles or products of their decay constitute a part
of cold dark matter) - a rather
strong assumption. Finally, even so, their amplitude (in sharp
contrast with adiabatic perturbations) does depend on the form of
the transition from inflation to the radiation-dominated FRW stage.
Nevertheless, isocurvature perturbations represent an interesting and
important object of investigation, especially because some candidates
for such inflatons that might survive from the inflationary era up to
the present
time are already known - dilatons in the Brans-Dicke theory and
superstring induced theories (see ~\cite{gasp} for the latter), axions
in ``natural'' inflation ~\cite{adams}, etc.

So, in the present paper we consider isocurvature perturbations in the
simplest case when $N$ scalar fields have arbitrary self-interaction
potentials but interact mutually through gravity only, i.e. the
interaction potential $V(\phi_1,~...\phi_N)=\sum_1^N V_n(\phi_n)$.
The general quantitatively correct expression for adiabatic
perturbations generated in this model was obtained in ~\cite{star85}.
First, we find the general solution for the behaviour of long-wave
isocurvature perturbations during the multiple inflationary stage and then
determine the correct coefficients in it by exact matching to vacuum
quantum fluctuations of the scalar fields in the approximately
de Sitter background during the inflationary stage (Sec. 2). After
that, in Sec. 3, we investigate
some cases where it is possible to match this solution to the
radiation-dominated FRW model with a small admixture of non-thermalized
massive particles (``cold dark matter''). The most interesting case
with respect to cosmological applications turns out to be the double
inflationary model with two massive inflatons, the heavier one
remaining uncoupled from usual matter after inflation. The
isocurvature perturbation spectrum in this model has a maximum on
small scales whose value may be rather large.

\section{Behaviour of perturbations during a multiple inflationary stage}

We consider the following Lagrangian density describing gravity plus $N$
scalar fields
\begin{equation}
L=-{R\over {16\pi G}}+\sum_{j=1}^N \Bigl ({1\over
2}\phi_{j,\mu}\phi_j^{,\mu}- V_j(\phi_j) \Bigr )
\end{equation}
where $\mu=0,..,3,~c=\hbar=1$ and the Landau-Lifshitz sign
conventions are used.
Note that the $N$ scalar fields interact only gravitationally.
The space-time metric has the form
\begin{equation}
ds^2=dt^2-a^2(t)\delta_{mn}dx^m dx^n, \qquad\ m,n=1,2,3.
\end{equation}
Spatial curvature may always be neglected because it becomes
vanishingly small after the first few e-folds of inflation.
The homogeneous background is treated classically, it is determined by the
scale factor $a(t)$ and the $N$ scalar fields $\phi_j(t)$. Their
equation of motion is given by
\begin{eqnarray}
H^2=\sum_{j=1}^N {{8\pi G}\over 3}
\Bigl ({\dot \phi_j^2\over 2}+ V_j(\phi_j)\Bigr ), \\
\ddot \phi_j+3H\dot \phi_j+V'(\phi_j)=0,~~~~ \qquad \ j=1,..,N \label{eq}
\end{eqnarray}
where a dot denotes a derivative with respect to $t$ while a prime
stands for a derivative with respect to $\phi_j$.
 From (\ref{eq}), we get the useful equation
\begin{equation}
\dot H=-4\pi G \sum_{j=1}^N {\dot \phi_j}^2.
\end{equation}
In these models, therefore, $H$ always decreases with time.
\par
Let us turn now to the inhomogeneous perturbations.
We consider a perturbed FRW background whose metric, in the
longitudinal gauge, is given by
\begin{equation}
ds^2=(1+2\Phi)dt^2-a^2(t)(1-2\Psi)\delta_{mn}dx^mdx^n.
\end{equation}
We get from the perturbed Einstein equations ($exp(i{\bf {kr}})$ spatial
dependence is assumed and the Fourier transform convention is
$\Phi({\bf k})\equiv (2\pi)^{-3/2}\int
\Phi({\bf r})e^{-i{\bf kr}}d^{3}{\bf k}$)
\begin{eqnarray}
\Phi & = & \Psi,\label{eq1}\\
{\dot \Phi}+H\Phi & = & 4\pi G \sum_{j=1}^N {\dot \phi_j}
\delta\phi_j ,\label{eq2}\\
\delta \ddot{\phi_j} + 3H\delta \dot{\phi_j} + \Bigl({{k^2}\over{a^2}}+
V_j''\Bigr )\delta \phi_j & = & 4\dot{\phi_j} \dot{\Phi}-2V_j'\Phi, \qquad\
j=1,..,N.\label{eq3}
\end{eqnarray}

We see that when we have more than one scalar field,
the dynamics of the perturbed system cannot be described by just one
equation for the master quantity $\Phi$. It is remarkable
that without solving the system (\ref{eq1}-\ref{eq3})
we can immediately write its two exact solutions desribing
the adiabatic modes in the formal limit $k\to 0$:
\begin{eqnarray}
\Phi & = & C_1\left(1-{H\over a}\int_0^t adt'\right)+C_2{H\over a},
\label{ad1}\\
{\delta \phi_j\over \dot \phi_j} & = & {1\over a}\left(C_1\int_0^t adt'
-C_2\right), ~~~~\qquad\ j=1,..,N\label{ad2}
\end{eqnarray}
where  $a(t),~\phi_j(t)$ satisfy the exact background equations (3,4)
and  $C_1$ and $C_2$ may still depend on ${\bf k}$. The term with
$C_1$ is the growing adiabatic mode, the term with $C_2$ is
the decaying adiabatic one. The existence of these exact solutions
directly follows from the observation made in ~\cite{star82} (see also
the detailed explanation in ~\cite{david}) that there always exists a
solution for scalar perturbations in the flat (${\cal K}=0$) FRW
Universe which has the following asymptotic behaviour
in the synchronous gauge in the limit $k\to 0$ in terms of the
Lifshitz variables: $\mu ({\bf k})= 3h({\bf k}),~\lambda({\bf k}) =0,~
\delta \phi_j({\bf k})=0$
(with no dependence
on $t$) {\it irrespective of the structure and the properties of the
energy-momentum tensor of matter}. Knowledge of the solutions (10,11)
is not, however, sufficient to find
the amplitude of generated perturbations if the number of scalar fields
$N>1$, in that case we have to integrate the system (7-9) completely in
the limit $k \to 0$ at the inflationary stage.

Let us now consider a multiple inflationary stage with $\tilde N$
background scalar fields being in the slow-rolling regime
$(\tilde N\le N)$, $\tilde N$ may depend on time. The energy
density of all other scalar fields not being in the slow-rolling regime
decreases exponentially with time and soon becomes negligible, thus, these
fields should be simply omitted from the background equation (3).
Then (3,4) simplify to
\begin{eqnarray}
H^2 & = & \sum_{j=1}^{\tilde N} {{8\pi G}\over 3} V_j(\phi_j), \\
3H\dot \phi_j+V'(\phi_j) & = & 0,~~~~ \qquad \ j=1,..,\tilde N .
\end{eqnarray}
Now, for $k\ll aH$, the system (7-9) can be solved in a way completely
analogous to ~\cite{star85,david}. First, its solutions corresponding
to growing adiabatic and non-decreasing isocurvature modes weakly
depends on time, so
for them Eqs. (8,9) take the form:
\begin{eqnarray}
\Phi & = & {4\pi G \over H}\sum_{j=1}^{\tilde N} \dot{\phi_j}
\delta \phi_j,\\
3H\delta \dot{\phi_j} + V_j''\delta \phi_j & = & -2V_j'\Phi,
\qquad\ j=1,..,\tilde N.
\end{eqnarray}
The general solution is
\begin{eqnarray}
\Phi & = & -C_1{{\dot H}\over {H^2}}-H{d\over {dt}}
\Bigl (\frac{\sum_j d_j V_j}{\sum_j V_j} \Bigr ),\label{sol1}\\
{{\delta \phi_i}\over {\dot \phi_i}} & = & {{C_1}\over H}-2H\Bigl
(\frac{\sum_j d_jV_j}{\sum_j V_j}-d_i\Bigr ),~~~~
i,j = 1,..,\tilde N. \label{sol2}
\end {eqnarray}
Here  $C_1$ and $d_j$ are integration constants, only
$\tilde N-1$
out of the $\tilde N$ coefficients $d_j$ are linearly independent,
and we will further use this freedom to add a constant term to them.
The background quantities $H(t), \phi_j(t)$ are exact solutions
of Eqs. (12,13). The mode with the coefficient
$C_1$ is the growing adiabatic mode as can be seen from the comparison
with (10,11), the other $\tilde N-1$ modes are the non-decreasing
isocurvature modes.

The expression for the decaying adiabatic mode immediately follows
from the general expressions (10,11) which take the following form at
the inflationary stage:
\begin{equation}
\Phi =C_2{H\over a},~~~~{\delta \phi_j\over \dot \phi_j}= -{C_2\over a}
\end{equation}
for all scalar fields (including those which are not in the
slow-rolling regime). The expression for decaying isocurvature
modes of slowly-rolling scalar fields may be found, similarly
to~\cite{david}, by assuming that all
quantities in Eqs. (7-9) are proportional to $a^{-3}(t)$ multiplied by
slowly varying functions of $t$ (note that the approximate form
(14,15) of these equations cannot be used now). The answer is
\begin{equation}
\Phi =\Psi =0,~~~\delta \phi_j={\tilde d_j\over \dot \phi_j H^2a^3},~~~
\sum_j \tilde d_j=0, ~~~j=1,..,\tilde N
\end{equation}
which may be easily verified by direct substitution using (12,13).
Finally, all other $2(N-\tilde N)$ scalar modes connected with
non-slowly-rolling scalar fields are decreasing isocurvature ones, too.
We shall consider them below in connection with matching to a
post-inflationary era.

Let us return to the most interesting non-decreasing modes.
Another quantity which is useful for their description is the
fractional comoving energy perturbation in each scalar field component
\begin{equation}
\Delta_j\equiv {\delta \varepsilon_j^{(c)} \over (\varepsilon +p)_j}=
{\dot \phi_j \delta \dot \phi_j + V'_j\delta \phi_j +
3H\dot \phi_j \delta \phi_j - \dot \phi_j^2 \Phi \over
(\varepsilon +p)_j}= {\partial \over \partial t}\left ({\delta \phi_j
\over \dot \phi_j}\right) - \Phi
\end{equation}
where $\delta \varepsilon_j^{(c)}$ coincides with Bardeen's gauge-invariant
quantity $\epsilon_m$ times the background energy density in the case of one
scalar field (see ~\cite{bardeen}) and relation (7) is used.
Note also the following consequence of Eqs. (7-9) which is actually
the Newton-Poisson equation in the cosmological case:
\begin{equation}
k^2 \Phi = -4\pi Ga^2\sum_{j=1}^N \delta \varepsilon_j^{(c)}~.
\end{equation}

In the long-wave limit $k\to 0$, the substitution of expressions
(10,11) into (20) gives $0$. This means that the small-$k$ expansion of
$\Delta_j$ contains an additional $k^2$ multiplier in the case of both
adiabatic modes, i.e., $|\Delta_j|\ll |\Phi|$ for them in this limit.
On the other hand, $|\Delta_j|$ can be of the order of, and even much
bigger
than, $|\Phi|$ for isocurvature modes though the total comoving density
perturbation $\sum_{j=1}^N \delta \varepsilon_j^{(c)}$ still contains the
additional $k^2$ multiplier compared to $\Phi$ as follows from (21).
Substituting the expressions (16,17) valid during the multiple inflationary
stage into (20) we get:
\begin{equation}
\Delta_i = 2d_i\dot H + 8\pi G\sum_j d_j \dot \phi_j^2,~~~j=1,..,\tilde N.
\end{equation}

The next step is to determine the coefficients $C_1,~d_j$ from
amplitudes of quantum fluctuations of scalar fields generated during
the inflationary stage. First, we invert (16,17) to obtain:
\begin{eqnarray}
C_1 & = &{8\pi G\over 3H}\sum_j{V_j\over \dot \phi_j}\delta \phi_j
= - 8\pi G \sum_j {V_j\over V_j^{\prime}}\delta \phi_j~, \\
d_i & = &{\delta \phi_i \over 2H\dot{\phi_i}}-{C_1\over 2H^2}+
{\sum_j d_j V_j\over \sum_j V_j}~, ~~~i,j = 1,..,\tilde N.\label{di}
\end{eqnarray}
Further, using the above-mentioned possibility to add the same
constant to all $d_j$, we omit the last two terms in Eq. (24).
Then all $\delta \phi_j$ in the r.h.s. of (23,24) have to be matched
with quantum fluctuations of the scalar fields generated during
the inflationary stage (we remind that this is a genuine
quantum-gravitational effect). For all scalar fields being in the
slow-rolling regime, $|m_{j,eff}^2|\equiv |V''_j|\ll H^2$, therefore,
all mass- and $\Phi$- dependent terms in Eqs. (8-9) may be neglected
for $k\ge aH$, and even in the region $k<aH$ but $(t-t_k)H(t_k)\ll
H^2(t_k)/|\dot H(t_k)|$ where  $t_k$ is the time when a mode $k$
crosses the Hubble radius during the inflationary stage, i.e.,
$k=a(t_k)H(t_k)$ (it is in the latter region where the exact matching
is performed). Then the $\delta \phi_j$'s behave like massless
uncoupled scalar fields in the de Sitter background. The standard
quantization gives the well-known result (see e.g.~\cite{star82}): for
$k\ll aH$, the Fourier components of the fields are time independent
("frozen") and may be represented in the following form:
\begin{equation}
\delta \phi_j({\bf k})={{H(t_k)}\over {\sqrt {2k^3}}}e_j({\bf k})\label{dS}
\end{equation}
where $e_j({\bf k})$ are classical
stochastic Gaussian quantities with vanishing average values
$< e_j({\bf k}) >=0$ and the correlation matrix
$< e_j({\bf k})e_{j^{\prime}}^{*}({\bf k'}) >=\delta_{jj^{\prime}}
\delta^{(3)}
({\bf k}-{\bf k'})$. Note however that in the case of multiple inflation
there may exist small effects ~\cite{gw} for which the approximation
(25) is not sufficient and one has to take into account a small
quantum correction
to it reflecting the fact that the generated fluctuations are in a
squeezed pure quantum state with a large but finite squeezing
parameter $r$ (the limit $r\to \infty$ is completely equivalent to (25)).
As for scalar fields with large effective masses which are not in a
slow-rolling regime, their fluctuations are negligible (apart from the
case when they experience a non-equilibrium first-order phase
transition during inflation which we do not consider here).

By substituting (25) into (23,24), we get finally
(we denote by $C_1^2(k)$ the power spectrum of the stochastic
quantity $C_1({\bf k})$ and use a similar notation for all stochastic
variables, $\langle f({\bf k}) f^*({\bf k'})\rangle=f^2(k)\delta^{(3)}
({\bf k}-{\bf k'})$):
\begin{eqnarray}
C_1({\bf k}) & = & -{8\pi GH\over \sqrt{2k^3}}\sum_j {V_j\over
V_j^{\prime}}e_j~,~~~~C_1^2(k) =  {32\pi^2G^2H^2\over k^3}\sum_j{V_j^2
\over V_j^{\prime 2}}~, \\
d_i({\bf k}) & = & -{3H\over 2\sqrt{2k^3}V_i^{\prime}}e_i~, ~~~~
d_i^2(k) =  {9H^2\over 8k^3V_i^{'2}}~,~~~~i,j=1,..,\tilde N
\end{eqnarray}
where all the time-dependent quantities in the r.h.s. are taken at $t=t_k$.
The result for $C_1^2(k)$ coincides with that previously obtained
in ~\cite{star85}. In the case of two scalar fields ($j=1,2$), we reproduce
the results of ref.~\cite{david} where the notation $C_3\equiv d_1-d_2$
was used. The expressions (16,17,26,27) are the main results of this section.

\section{Matching to a post-inflationary era}

As was mentioned in the introduction, in the case of isocurvature
modes we don't have general expressions like (10,11) for adiabatic
modes, hence the post-inflationary behaviour of isocurvature
perturbations is
not universal and depends on additional assumptions. In particular,
there could be no such perturbations at all soon after the end of an
inflationary stage. Therefore we will consider further a number
of specific models in
which they may be present even nowadays. The most natural way to achieve
it is to assume that one of the inflaton scalar fields remains uncoupled
from usual matter (baryons, photons, etc.) all the time since the end of
the multiple inflationary stage up to the present moment, and that its
particles or products of their decay (still uncoupled from usual matter)
constitute today a part of the cold dark matter with a dust-like equation
of state ($p\ll \varepsilon$).

The non-decreasing mode of isocurvature fluctuations in a
system of two uncoupled components consisting of dust-like matter on
one hand, and radiation coupled to baryons on the other hand, can be
characterized by a fractional
comoving energy density perturbation in the dust-like component
\begin{equation}
\delta_m\equiv {\delta \varepsilon_m^{(c)}\over \varepsilon_m}=
\delta_i{\Omega_i\over \Omega_m}~,~~~~\delta_i\equiv {\delta
\varepsilon_i^{(c)}\over \varepsilon_i}\approx \Delta_i~,
\end{equation}
that remains constant during the radiation-dominated era
(see, e.g., \cite{mfb}-\cite{ll} for reviews).
Here $\Omega_i$ is the present-day density (in terms of the critical one)
of that part of cold dark matter which is the relic from the
inflationary era while $\Omega_m$ refers to all the cold dark
matter ($\Omega_i\le \Omega_m$). Note that we have for the number density
perturbation, ${\delta n_i\over n_i}=\delta_i$ where $n_i$ is the
number density of relics.
$\Omega_{tot}=\Omega_m+\Omega_{bar}=1$ with great accuracy for
cosmological models having an inflationary stage (the energy density
of the cosmological term, if non-zero, should be added to
$\Omega_{tot}$, too). After the transition to the matter-dominated stage
at redshifts $z\approx 10^4$, this mode produces a growing adiabatic
mode of fluctuations which evolves, as usually, $\propto a(t)$ afterwards.

Therefore, we have to relate $\delta_i$ at the radiation-dominated stage
with $\Delta_i$ at the inflationary stage, as given in Eq. (22).
We further specialize to the case of two inflaton scalar fields
and replace the subscripts 1, resp. 2 by $h$ (heavy),
resp. $l$ (light). Then Eqs. (16,17,22) take the following form which
generalizes the results of ref.~\cite{david}:
\begin{eqnarray}
\Phi(t) & = & -C_1({\bf k}){{\dot H}\over {H^2}}+{C_3({\bf k})\over 3}
\frac {V_l V_h^{\prime 2}- V_h V_l^{\prime 2}}{(V_h+V_l)^2}~, \\
{\delta \phi_h\over \dot \phi_h}(t) & = & {C_1({\bf k})\over H}+
2C_3({\bf k})~{H V_l\over V_h+V_l}~, \\
{\delta \phi_l\over \dot \phi_l}(t) & = & {C_1({\bf k})\over H}-
2C_3({\bf k})~{H V_h\over V_h+V_l}~, \\
\Delta_h (t)& = & -{C_3({\bf k})\over 3}~{V_l^{'2}\over V_h+V_l}~, \\
\Delta_l (t)& = & {C_3({\bf k})\over 3}~{V_h^{'2}\over V_h+V_l}
\end{eqnarray}
where $C_1$ and $C_3 = d_h-d_l$ are given in Eqs. (26,27). Two
essentially different cases may take place
which we call the cases of heavy relics and light relics.
\bigskip

1. Heavy relics. \\

\noindent
This case arises when the inflaton field that remains uncoupled
from usual matter after inflation has an effective mass larger than
$H$ at the end of inflation. Then this ``heavy'' scalar field $\phi_h$
is in the slow-rolling regime in the first part of inflation, but it
goes out of this regime when $H$ becomes less than the effective mass
during inflation. Somewhat earlier, its energy density becomes much
smaller than the total one. Let us take the simplest case where the
effective
mass is constant, so that the potential is $V_h=m_h^2 \phi_h^2/2$,
and $G\phi_h^2\gg 1$ at the early stages of inflation (for the field to be
in the slow-rolling regime initially). Note for completeness that it is
not possible to realize such a scenario for a steeper power-law
potential $V_h$, in particular for $V_h=\lambda \phi_h^4/4$, because
then the effective mass remains smaller than $H$ till the end of inflation.

If $\varepsilon_h\ll \varepsilon_{tot}$, then irrespective of the
fact whether the field $\phi_h$ is in the slow-rolling regime or not, the
right-hand side of Eq.(9) may be neglected for isocurvature modes.
Now we need to solve this equation in the limit $k^2=0$. Note that
then the left-hand sides of Eqs.(9) and (4) coincide in the case of
a massive scalar field without self-interaction. So, one of the
solutions is
$\delta \phi_h \propto \phi_h(t)$ where $\phi_h(t)$ is the exact solution
of Eq.(4) in a background driven by the other scalar field through
Eq.(3). The other linearly independent solution can be found from
the Wronskian condition but we don't need it here because,
if $\phi_h$ is still in the slow-rolling regime,
Eq.(30) may be applied which reads $\delta \phi_h = 2C_3H\dot \phi_h =
-2C_3m_h^2\phi_h/3$ for $V_h\ll V_l$. Now we use the constancy of
the quantity $\delta \phi_h(t)/\phi_h(t)$ during a transition from
inflation to the radiation-dominated stage. For $m_ht\gg 1$ at the
latter stage, the heavy field is in the WKB regime of oscillations
with the frequency $m_h$ (see, e.g., \cite{david} for exact expressions).
Averaging over the oscillations, we obtain
\begin{equation}
\delta_h = 2~{\delta \phi_h\over \phi_h}= -{4\over 3}m_h^2C_3({\bf k})~.
\end{equation}
Note that though Eq. (34) looks like Eq. (19.18) in~\cite{mfb}, it is
not exactly the same because it refers to a different quantity (a comoving
energy perturbation vs. an energy perturbation in the longitudinal
gauge (6)), and we apply it in a different regime (at the
radiation-dominated vs. the inflationary stage). From (27),
the amplitude of the fractional energy and number density perturbation
in the relic dust component during the radiation-dominated stage follows:
\begin{equation}
k^3\delta_h^2(k)= 2H^2(t_k)\left({1\over \phi_h^2}+
{m_h^4\over V_l^{'2}}\right)_{t_k}
\end{equation}
where the first term inside the brackets in the r.h.s. of the last
equation should be omittted if $H(t_k)<m_h$. The power spectrum has
a slope $n\approx -3$ similar to that of $\Phi^2(k)$ for adiabatic
perturbations.

To make a quantitative comparison between contributions of
isocurvature and adiabatic modes to effects observable today,
one should take into account that, due to the properties of the transfer
function for isocurvature fluctuations in the CDM+radiation model,
an isocurvature density fluctuation $\delta_m$ at the
radiation-dominated stage produces the same adiabatic mode after
transition to the matter-dominated stage ($a(t)\propto t^{2/3})$ as
the initial adiabatic mode with $\Phi=\delta_m/5$ for scales exceeding
by far
the present comoving scale corresponding to the cosmological horizon at
the moment of
matter-radiation equality $R_{eq}\approx 30h^{-1}Mpc$, $H_0\equiv
h\cdot 100~km~s^{-1}Mpc^{-1}$
($\Omega_{bar}$ is assumed to be small, too). For $kR_{eq}>1$,
the equivalent amplitude of $\Phi$ is even less. On the other hand,
isocurvature fluctuations produce $6$ times larger angular
temperature fluctuations $\Delta T/T$ in the CMB at angles $\theta >
30'$ for the same amplitude of long-wave density perturbations at the
matter-dominated stage, i.e., $\Delta T/T= 2\delta_m/5$ vs.
$\Delta T/T=\Phi/3$ for adiabatic perturbations~\cite{ss},~\cite{eb},
{}~\cite{ks}
(see also~\cite{mfb}-\cite{ll} for reviews). Due to the latter
reason, it has
been long known that isocurvature fluctuations with a flat ($n=-3$)
initial spectrum cannot be responsible for the observed large-scale
structure and $\Delta T/T$ fluctuations in the Universe.

For a power-law $V_l$ with the last part of inflation driven by the
light scalar field, $V'_h > V'_l$ in the region $V_h\sim V_l$ where
the transition from heavy to light scalar field
domination of the total energy takes place. Then the second term inside
the brackets in (35) is the dominant one, while still much smaller than
$\Phi^2(k)=9C_1^2(k)/25$. Therefore isocurvature
fluctuations, if present, are less than adiabadic ones in double
inflationary models in the region around the break in the perturbation
spectrum due to a transition between the two phases of
inflation, and their possible presence changes
nothing regarding the confrontation of these models with observational
data (see, e.g.~\cite{ppst}). But on much smaller scales, when the first
term inside the brackets in (35) is dominant, isocurvature perturbations
become much larger. In that case
\begin{equation}
k^3\delta_h^2(k)={2H^2\over \phi_h^2}(t_k)~.
\end{equation}
Alternatively, this result may be
obtained very simply by considering the heavy field as a test field in the
de Sitter background and using the expressions (25,34). The spectrum (36)
grows with $k$ (because $\phi_h$ quickly decreases with $t$) until the
point $H(t_k)\sim m_h$ is reached, after that it falls abruptly.

If, e.g., $V_l={1\over 2}m_l^2\phi_l^2$ and $m_l\ll m_h$, then, using Eqs.
(2.10-2.15) of ref.~\cite{david} and Eq. (35), we get
\begin{equation}
k^3\delta_h^2(k)= {8\pi Gm_l^2\over 3}\left( \left({s_0
\over s_0 - \ln {k\over k_b}}\right)^{m_h^2/m_l^2} + \left({m_h\over
m_l}\right)^4\right),~~~k\gg k_b
\end{equation}
where $k_b$ is the location of the break and $s_0\gg 1$ is the number of
e-folds during the second phase of inflation driven by the light
scalar field ($s_0\approx 60$ to account for observational data,
see~\cite{ppst}). The expression (37) is derived under the approximation
$m_h^2/m_l^2 < s_0$ which corresponds to the absence of a power-law
intermediate stage between the two phases of inflation (double inflation
without break, according to the terminology of~\cite{david}), a more
suitable
condition perhaps is the absence of oscillations or a smooth transition
in the spectrum, which will be the case for
$m_h/m_l<15$~\cite{david94} or $m_h^2/m_l^2 < 4s_0$. In the
opposite case of double inflation with a break, there is no growth of
isocurvature fluctuations at small scales. Note that the effect of
growth in the isocurvature perturbation spectrum was previously
noticed from numerical calculations for a similar model in~\cite{lev88}.

For not too small scales when $\ln (k/k_b)\ll s_0$, the first term in
the spectrum (37)
is power-law like with a small exponent: $k^3\delta_h^2(k)\propto
(k/k_b)^{\alpha},~\alpha = m_h^2/m_l^2s_0$. The
perturbations reach their maximum on short scales for which $s_0 - \ln (k/k_b)
\sim m_h^2/m_l^2$, due to the disappearance of the first
term in (37) for $H(t_k)<m_h$, its value being
\begin{equation}
\left(k^3\delta_h^2(k)\right)_{max}\sim Gm_l^2 \left({m_l^2s_0
\over m_h^2}\right)^{m_h^2/m_l^2}~.
\end{equation}
It is interesting that for $\sqrt{G}m_l\sim 10^{-6},~s_0\sim 60$
the maximal value of the quantity (38) as a function of $m_h/m_l$, though
still smaller than unity, is not far from it (it is reached for
$m_h/m_l \approx 5$). Hence, such a model can be used to produce a
significant
number of primordial black holes with rather small masses (for a
review of observational upper limits on the number density of PBHs see,
e.g.,~\cite{npsz}-\cite{carr}). Then, however, it cannot explain the
observed large-scale structure and $\Delta T/T$ fluctuations in the
Universe because $m_h/m_l$ is required to be $\approx 12-14$
(and certainly more than $8$) for this aim, see~\cite{ppst}.
\bigskip

2. Light relics \\

\noindent
In this case, $m^2_{eff}\ll H^2$ for one of the inflaton scalar fields
during the whole inflation. To avoid this ``light'' field to be
dominating during the last part of inflation and after its end, we
have to assume that its energy density $\varepsilon_l\ll
\varepsilon_{tot}$ during inflation, too. Then we have from Eq. (31):
\begin{equation}
\delta \phi_l = -2C_3H\dot \phi_l ={2\over3}C_3V'_l~.
\end{equation}
If $C_3$ from Eq. (27) is substituted into this expression and it is
assumed that $V'_l \ll V'_h$, the standard  expression (25) for
fluctuations of a test scalar field on a de Sitter background
arises once more. Let $V_l={1\over 2}m_l^2\phi_l^2$, then we may
repeat the derivation made in the previous subsection to get the
expression (36) (with the index ``h'' changed to ``l'') for the
fractional density perturbation in the light relic component.
Now $\phi_l$ is practically constant during inflation (and less than
$G^{-1/2}$ to avoid a second inflationary phase), so the
spectrum is falling with $k$. Isocurvature perturbations may be larger
than adiabatic ones if $\phi_l$ is small enough, but this does not
lead to interesting cosmological models for a smooth potential $V_h$
satisfying the slow-rolling conditions due to
the reason mentioned in the previous subsection in connection
with the $n=-3$ initial isocurvature perturbation spectrum.
\bigskip

3. Intermediate relics \\

\noindent
Let us briefly consider the case of an inflaton field which remains
uncoupled from usual matter after inflation, dominates during the first
phase of inflation (so we call it ``heavy'') and has the quartic
potential $V_h=\lambda_h \phi_h^4/4$. During the last part of inflation,
$s<s_0$, where $s$ is the number of e-folds measured from the end of
inflation and $s_0\gg 1$ is the moment when $V_h=V_l$,
$\varepsilon_h\ll \varepsilon_{tot}$, however $m^2_{h,eff}\equiv
3\lambda_h\phi_h^2 \ll H^2$ (and less than the effective mass of the
other scalar field, too). So, this initially ``heavy'' inflaton becomes
``light'' in the last part of inflation. That is why we call this case
the intermediate one.

Then, for $s\ll s_0$, $\delta \phi_h = 2C_3H\dot \phi_h =
-2C_3\lambda_h\phi_h^3(t)/3$ as in the case of massive relics. Here
the quantity $\delta \phi_h/\phi_h$ is not constant during the last
period of inflation and the transition to the radiation-dominated stage.
Therefore, an exact matching (as it was done in the subsection 1) is not
possible, but we may make a matching by order of magnitude using
the fact that $\delta_h$ at the radiation-dominated stage is of the
order of $\delta \phi_h/\phi_h$ at the end of inflation.
Using (30,34), we get: $\delta_h^2(k)=const \cdot
C_3^2(k)m^4_{h,eff}(t_f)$ where $t_f$ is the moment when inflation
ends and $const = {\cal O}(1)$.

\bigskip

a) $V_l=m_l^2\phi_l^2/2~, ~~\lambda_h \gg Gm_l^2~.$ \\
\noindent
Then $\phi_h^2 =m_l^2/\lambda_h\ln (s_0/s)$ during the last period of
inflation. Therefore, $m^2_{h,eff}(t_f)=3m^2_l/\ln s_0$, and we arrive
at the following result
\begin{equation}
k^3\delta_h^2(k)=const \cdot H^2(t_k)\Biggl (\frac{1}{\lambda_h^2 \phi_h^6}
+\frac{1}{m_l^4 \phi_l^2}\Biggr )_{t_k}~\frac{m_l^4}{\ln^2 s_0}
\end{equation}
where $const = {\cal O}(1)$. In particular, for $s\ll s_0$,
\begin{equation}
k^3\delta_h^2(k)=const \cdot {\lambda_h s(k)\ln^3 \left(s_0/s(k)\right)
\over\ln^2 s_0}
\end{equation}
where $s(k)\equiv s(t_k)=\ln (k_f/k)\gg 1$ and $k_f=a(t_f)H(t_f)$.
The spectrum is approximately flat. Though it has a smooth maximum
at $s=s_0e^{-3}$, this maximum is not as strongly pronounced as
in the case of massive heavy relics. The amplitude grows
$\propto (k/k_b)^{1.5}$ for $s_0-s\ll s_0$ where $k_b$ is the
inverse comoving scale corresponding to the first horizon crossing
at the moment $s_0$.
\bigskip

b) $V_l=\lambda_l\phi_l^4/4~,~~\lambda_h\gg \lambda_l~.$ \\
\noindent
Now $\phi_h^2=\lambda_l \phi_l^2/2\lambda_h\ln (s_0/s)$ during the second
phase of inflation. Thus, $m^2_{h,eff}(t_f)\sim \lambda_l/G\ln s_0$ and
\begin{equation}
k^3\delta_h^2(k)=const \cdot H^2(t_k)\Biggl (\frac{1}{\lambda_h^2
\phi_h^6}+\frac{1}{\lambda_l^2 \phi_l^6}\Biggr )_{t_k}~\frac{\lambda_l^2}
{G^2\ln^2 s_0}
\end{equation}
where $const = {\cal O}(1)$. In particular, for $s\ll s_0$,
\begin{equation}
k^3\delta_h^2(k)=const \cdot {\lambda_h\ln^3 \left(s_0/s(k)\right)\over s(k)
\ln^2 s_0}~.
\end{equation}
The spectrum is approximately flat and grows slightly towards large $k$'s.
Once more, its amplitude grows $\propto (k/k_b)^{1.5}$ for
$s_0-s\ll s_0$.
\bigskip

\noindent
{\bf Acknowledgements}

\noindent
A.~S. is grateful to Profs. Y. Nagaoka and J. Yokoyama for their
hospitality at the Yukawa Institute for Theoretical Physics, Kyoto
University. The financial support for research work
of A.~S. in Russia was provided by the Russian Foundation for
Basic Research, Project Code 93-02-3631, and by Russian Research
Project ``Cosmomicrophysics''. D.~P. would like to thank Prof. J. Yokoyama
for hospitality at the Yukawa Institute and financial support.


\begin{thebibliography}{99}
\bibitem{star85} A.~A.~Starobinsky, JETP Lett. {\bf 42}, 152 (1985)
\bibitem{linde}A.~Linde, Rep. Prog. Phys. {\bf 47}, 925 (1984); {\it
Particle physics and inflationary cosmology}, (Harwood, New-York,
1990); E.~Kolb, M.~Turner,{\it The Early Universe}, (Addison-Wesley,
1990)
\bibitem{linde85}L.~A.~Kofman, A.~D.~Linde, A.~A.~Starobinsky,
Phys. Lett. B {\bf 157}, 361 (1985)
\bibitem{lev87}L.~A.~Kofman, A.~D.~Linde, Nucl. Phys. B {\bf 282},
555 (1987)
\bibitem{silk}J.~Silk, M.~S.~Turner, Phys. Rev. D {\bf 35}, 419 (1987)
\bibitem{lev88}L.~A.~Kofman, D.~Yu.~Pogosyan, Phys. Lett. B {\bf 214},
508 (1988)
\bibitem{bond}D.~S.~Salopek, J.~R.~Bond, J.~M.~Bardeen, Phys. Rev. D
{\bf 40}, 1753 (1989)
\bibitem{stefan}S.~Gottl\"{o}ber,
V.~M\"{u}ller, A.~A.~Starobinsky, Phys. Rev. D {\bf 43}, 2510 (1991)
\bibitem{david}D.~Polarski, A.~A.~Starobinsky, Nucl. Phys. B {\bf
385}, 623 (1992)
\bibitem{la89}D.~La, P.~J.~Steinhardt, Phys. Rev. Lett. {\bf 62}, 376
(1989)
\bibitem{muc}S.~Gottl\"{o}ber, J.~P.~M\"{u}cket,
A.~A.~Starobinsky, astro-ph/9309049, Ap. J. (1994) in press
\bibitem{ppst}P.~Peter, D.~Polarski, A.~A.~Starobinsky, preprint
YITP/U-94-06, DAMTP-R94/20, astro-ph/9403037 (1994), subm. for publ.
to Phys. Rev. D
\bibitem{star79}A.~A.~Starobinsky, JETP Lett. {\bf 30}, 682 (1979)
\bibitem{lin84}A.~D.~Linde, JETP Lett. {\bf 40}, 1333 (1984)
\bibitem{gasp}M.~Gasperini, G.~Veneziano, preprint CERN-TH.7178/94,
gr-qc/9403031 (1994)
\bibitem{adams}K.~Freese, J.~A.~Frieman, A.~V.~Olinto,
Phys. Rev. Lett. {\bf 65}, 3233 (1990)
\bibitem{bardeen}J.~M.~Bardeen, Phys. Rev. D {\bf 22}, 1882 (1980)
\bibitem{star82}A.~A.~Starobinsky, Phys. Lett. B {\bf 117}, 175 (1982)
\bibitem{gw}D.~Polarski, A.~A.~Starobinsky, preprint YITP/U-94-07 (1994)
\bibitem{mfb}V.~F.~Mukhanov, H.~A.~Feldman, R.~H.~Brandenberger,
Phys. Rep. {\bf 215}, 203 (1992)
\bibitem{ll}A.~R.~Liddle, D.~H.~Lyth, Phys. Rep. {\bf 231}, 1 (1993)
\bibitem{ss}A.~A.~Starobinsky, V.~Sahni, in: Proc. 6th Soviet
Gravitation Conf., ed. V.~N.~Ponomarev (MGPI Press, Moscow, 1984), p.
77; V.~Sahni, Ph.D. (Candidate) thesis, Moscow University, 1984;
A.~A.~Starobinsky, Sov. Astron. Lett. {\bf 14}, 166 (1988)
\bibitem{eb}G.~Efstathiou, J.~R.~Bond, Mon. Not. R. Astron. Soc. {\bf
218}, 103 (1986)
\bibitem{ks}H.~Kodama, M.~Sasaki, Int. J. Mod. Phys. {\bf A2}, 491 (1987)
\bibitem{david94}D.~Polarski, Phys. Rev. D, to be published (1994)
\bibitem{npsz}I.~D.~Novikov, A.~G.~Polnarev, A.~A.~Starobinsky,
Ya.~B.~Zeldovich, Astron. Astroph. {\bf 80}, 104 (1979)
\bibitem{carr}B.~J.~Carr, J.~E.~Lidsey, Phys. Rev. D {\bf 48}, 543 (1993)
\end{thebibliography}
\end{document}